\documentstyle[preprint,aps,floats,epsfig,subeqn,amssymb]{revtex}

\newcommand{\be}{\begin{equation}}
\newcommand{\ee}{\end{equation}}
\newcommand{\bea}{\begin{eqnarray}}
\newcommand{\eea}{\end{eqnarray}}
\newcommand{\bml}{\begin{mathletters}}
\newcommand{\eml}{\end{mathletters}}

\newcounter{fixy}
\begin{document}
\newenvironment{fixy}[1]{\setcounter{figure}{#1}}
{\addtocounter{fixy}{1}}
\renewcommand{\thefixy}{\arabic{fixy}}
\renewcommand{\thefigure}{\thefixy\alph{figure}}
\setcounter{fixy}{1}

\tighten

\preprint{DCPT-02/29}
\draft




\title{Multiple algebraisations of an elliptic Calogero-Sutherland model}
\renewcommand{\thefootnote}{\fnsymbol{footnote}}
\author{ Yves Brihaye\footnote{Yves.Brihaye@umh.ac.be}}
\address{Facult\'e des Sciences, Universit\'e de Mons-Hainaut,
 B-7000 Mons, Belgium}
\author{Betti Hartmann\footnote{Betti.Hartmann@durham.ac.uk}}
\address{Department of Mathematical Sciences, University
of Durham, Durham DH1 3LE, U.K.}
\date{\today}
\setlength{\footnotesep}{0.5\footnotesep}

\maketitle
\begin{abstract}
Recently, G\'omez-Ullate et al. \cite{gome} 
have studied a particular $N$-particle quantum problem with 
an elliptic function potential supplemented by 
an external field. They have shown that the Hamiltonian
operator preserves
a finite dimensional space of functions and as such is
quasi exactly solvable (QES). 
In this paper we show that other types of invariant 
function spaces exist, which are in close relation to the 
algebraic properties of the elliptic functions.
Accordingly, series of new algebraic eigenfunctions can be 
constructed. 

\end{abstract}

\pacs{PACS numbers: 03.65.Fd}

\renewcommand{\thefootnote}{\arabic{footnote}}
\section{Introduction}

The first example of a non-trivial, integrable quantum
many-body Hamiltonian was found by Calogero \cite{calo}.
It describes a system of N particles in one dimension
interacting pairwise by means of an inverse square potential.
The similar model endowed with an inverse sine-square
potential is also integrable as shown by Sutherland
\cite{suth}. In fact these two potentials are particular
cases of a two parameter-family of potentials defined
by the Weierstass function \cite{calogero2,krichever}.
A detailed analysis of these
models generalising the Calogero-Sutherland (CS) quantum models
was reported in \cite{pere1}. Their classical counterparts
are  discussed e.g. in \cite{pere2}.

While integrable models (classical or quantum) can be studied
because of their mathematical interest, it became apparent in recent
years that the CS models can be applied to a large number of
fields of physics.  
These range from condensed matter (quantum Hall liquids, quantum spin chains, ..)
\cite{cmat}
to gauge theories \cite{gt}, soliton theory \cite{soliton} 
as well as recently to questions related to black holes and 
(Anti)-deSitter space \cite{gibbons,brigante}.
In particular it was shown in \cite{brigante}  
that the asymptotic dynamics of
$2$-dimensional gravity in Anti-deSitter and deSitter space
respectively 
can be described by a generalised two-body CS model. \

The property of a model to be integrable (i.e. to have a complete set
of commuting constants of motion) does not
necessarily imply that the spectrum and the eigenfunctions
of the corresponding Hamiltonian can be constructed explicitely.
The models which have this property are called {\it solvable}.
From the beginning the CS models were known to be solvable,
while further properties of their spectrum were obtained
only recently, see e.g. \cite{stanley,vinet}. However, the explicit
form of the spectrum is still missing as far as the full
Weierstrass-function potential is considered for generic values
of $N$.

A step forward in the construction of solvable N-body problems
interacting via a Weierstrass function was achieved in \cite{gome}.
The authors indeed showed that, when the Weierstrass potential
is supplemented by a suitable external potential, a finite
number of eigenvectors can be computed explicitely in terms of
special functions. Stated differently, the model is quasi
exactly solvable (QES) according to the definition of \cite{turbiner}.
In fact, the kind of interaction considered in \cite{gome} 
and in the present paper generalises a potential first
introduced in \cite{inom}.

Following the ideas of \cite{turbiner}, the QES property holds when the Hamiltonian
operator posseses a finite-dimensional invariant vector space of
 functions. 
Such a vector space was indeed constructed in \cite{gome}
for the Hamiltonian considered.
The purpose of this paper is to demonstrate that this Hamiltonian
possesses alternative invariant finite-dimensional vector spaces
of  functions.
The way these new vector spaces are constructed is very reminiscent
to the multiple algebraisations of the Lam\'e
equations (see e.g. \cite{bri1}), which occur due to the 
properties of the Jacobi elliptic functions.

 The Hamiltonian is presented in Sect. II. In this section we also give the 
transformation putting the Hamiltonian in a Lie-algebraic form which
reveals its QES property. 
The new invariant vector spaces are constructed in Sect. III 
and  the Hamiltonian is studied 
for particular values of the parameters.
The results are summarized in Sect. IV.

\section{An elliptic Calogero-Sutherland model}
The  quantum Hamiltonian proposed recently by G\'omez-Ullate et al. 
\cite{gome} is given by~:
\begin{equation}
\label{oper}
H_N({\bf x}) = -\sum^N_{k=1}{\partial^2\over{\partial x^2_k}} + V_N({\bf x}) 
\ , \ \ {\bf x}=(x_1,x_2,....,x_N) \ .
\end{equation}
It describes N particles on a line interacting  through the potential
\begin{equation}
V_N({\bf x}) = c_m\sum^N_{k=1} {\cal P}(x_k+i\beta)+4b(b-1)\sum^N_{k=1}
{\cal P}(2x_k)+a(a-1)
\sum^N_{j,k=1\atop{j\not=k}} [{\cal P}(x_j+x_k)+{\cal P}(x_j-x_k)] \ .
\label{potential}
\end{equation}
Here ${\cal P}(z) \equiv {\cal P}(z;g_2,g_3)$ denotes the Weierstrass function with invariants $g_2,g_3$. 
The constants $a$, $b$ are real and positive, $c_m$ is real. 
The term proportional to $c_m$ can be interpreted as the potential
of an external field.\

The Hamiltonian (\ref{oper}) was shown to admit an invariant, finite dimensional vector space of functions \cite{gome}. 
Restricting the operator to this vector space,
the eigenvalue equation $H_N \psi = E \psi$ is reduced to a matrix equation
and, accordingly, a finite number of eigenvectors can be determined algebraically.
Following the definition of \cite{turbiner} the operator $H_N$ is  called Quasi Exactly Solvable (QES).

To reveal this property $H_N$ has to be transformed appropriately. The authors
of Ref.\cite{gome} introduced the function (called  ``gauge factor'')
\begin{equation}
\mu({\bf x}) = \prod_{j<k} [{\cal P}(x_j+i\beta)-{\cal P}
(x_k+i\beta)]^a\prod_k[{\cal P}'(x_k+i\beta)]^b
\end{equation}
and the new variables 
\begin{equation}
z_k= {\cal P}(x_k+i\beta)\quad , \quad k=1,\cdots,N \ .
\end{equation}
Then a Hamiltonian $\bar H_N$ -spectrally equivalent to $H_N$-
is constructed  according to
 $\bar H_N({\bf z}) = \mu^{-1}({\bf z})H_N({\bf x})\mu({\bf z})$.
If  the coupling constant $c_m$ is chosen according to 
\begin{equation}
c_m = [2m+2a(N-1)+4b][2m+1+2a(N-1)+2b] \ , \ \ m \ \epsilon \  \mathbb{N}
\end{equation}
$\bar H_N({\bf z})$ preserves the finite dimensional polynomial space
\cite{tur}
\begin{equation}
{\cal M}_m=span\lbrace \tau_1^{l_1}\tau_2^{l_2}\cdots \tau^{l_N}_N\ ;\ \sum^N_{i=1}l_i\leq m\rbrace
\end{equation}
with the $k$-th elementary symmetric function
\begin{equation}
\label{tau}
\tau_k\equiv \sum_{i_1<i_2<\cdots <i_k} z_{i_1}z_{i_2}\cdots z_{i_k}\ ,\ 1\leq k\leq N \ . 
\end{equation}
A lengthy calculation leads to 
\begin{equation}
\label{hbar}
\bar H_N({\bf z}) = - \sum^N_{k=1}  p_k {\partial^2\over{\partial z^2_k}}-2a
\sum^N_{k,l=1\atop{k\not=l}} {p_k\over{z_k-z_l}} 
{\partial\over{\partial z_k}}
-(b+{1\over 2}) \sum^N_{k=1} p'_k{\partial\over{\partial z_k}} + \bar V_N({\bf z}) \ ,
\end{equation}
where $p_k \equiv p(z_k)$ and $p'_k \equiv p'(z_k)$ with 
\begin{equation}
p(z) =  4z^3-g_2z-g_3 \ , \ \  
p'(z) \equiv \frac{d p}{d z}= 12 z^2 -g_2 \ .
\end{equation}
In the following we will use the roots, $e_i$, $i=1,2,3$, of $p(z)$~:
\begin{equation}
 p(z) = 4(z-e_1)(z-e_2)(z-e_3) = 4z^3 - g_2 z - g_3  \  . 
\end{equation}
These numbers are equal to the values of the Weierstrass
function at its half-periods.

The potential $\bar V_N$ in (\ref{hbar}) is given by~:
\begin{equation}
\bar{V}_N({\bf z}) = m(12b+8a(N-1)+4m+2)\tau_1 \ , \ \ \tau_1=\sum^N_{k=1}z_k \ .
\end{equation}
The crucial observation is that
the Hamiltonian $\bar{H}_N({\bf z})$ can be written as a 
quadratic polynomial of the differential operators 
\begin{equation}
\label{lie}
{\cal D}_k=\frac{\partial}{\partial\tau_{k}} \ , \  \  
{\cal N}_{jk}=\tau_{j}\frac{\partial}{
\partial\tau_{k}} \ , \ \
{\cal U}_{k}=\tau_{k}(r - \sum_{i=1}^{N}
\tau_{i}\frac{\partial}{\partial\tau_{i}})  \ , \ \ j, k=1,2,..,N  \  
\end{equation}
with $r=m$. 
These operators form a representation of the Lie algebra $sl(N+1)$
for generic value of the real parameter $r$, for $r=m$ 
they preserve the vector space ${\cal M}_m$ and the representation
is finite dimensional.

Denoting by $\bar H_N^{(+)}$ the part of $\bar H_N$ which increases the degree of elements 
of ${\cal M}_m$, we find: 
\begin{equation}
\bar H_N^{(+)} = -4\tau_1({\cal N}-m)({\cal N}+m+{1\over 2} + 2a(N-1)+3b)
 \ , \ \ {\cal N} \equiv \sum_{k=1}^N {\cal N}_{kk} \ .
\label{pres}
\end{equation}
Obviously, the factor $({\cal N}-m)$
leads to the annihilation of all the monomials in ${\cal M}_m$ which have overall degree $m$.
Therefore, we find: 
$\bar H_N{\cal M}_m\subseteq {\cal M}_m$.
As a consequence, eigenvectors of $\bar H_N$ (and therefore also of $H_N$)
can be constructed in ${\cal M}_m$. In the following we will refer to this
property as to an ``algebraisation'' of $H_N$.

\section{Additional gauge  factors}
Inspired by the construction of the Lam\'e polynomials (see e.g. \cite{bri1}),
we introduce one further transformation of the Hamiltonian $H_N$~:
\begin{equation}
\bar H_N \rightarrow  \tilde H_N = \tilde\mu^{-1}\bar H_N\tilde \mu 
\end{equation}
with the gauge factor $\tilde \mu$ of the form~:
\begin{equation}
\label{factor}
\tilde \mu({\bf z}) =\prod_{k=1}^{N} (z_k-e_1)^{\nu_1}(z_k-e_2)^{\nu_2}(z_k-e_3)^{\nu_3} \ .
\end{equation}

The choice $\nu_1=\nu_2=\nu_3=0$ obviously corresponds to \cite{gome}. 
After a calculation, we find that  for each value of the form  
\begin{equation}
\nu_i = 0  \ \ {\rm or} \ \ 
\nu_i= {1\over 2}-b  \ ,\ \ \  i=1,2,3
\end{equation}
the Hamiltonian $\tilde H_N$ can be expressed as a quadratic combination
of the operators (\ref{lie}) with suitable values (depending
on the values of $\nu_i$'s) of the parameter $r$. We then found
eight gauge factors (\ref{factor}) leading to algebraisations of the initial
operator $H_N$. Let us now investigate the relations between $r$  and
the different parameters involved in the equations.

We find that the degree-increasing part, say $\tilde H_N^{(+)}$,
of  $\tilde H_N$ is given by:
\begin{equation}
\tilde H_N^{(+)} = -4\tau_1\left({\cal N}-(m+bn_f-{1\over 2}n_f)\right)
\left({\cal N}+m+2a(N-1)+(3-n_f)b+{1\over 2}(1+n_f)\right) \ .
\end{equation}
Here,
$n_f$ denotes the number of non-zero exponents $\nu_i$, $i=1,2,3$ in (\ref{factor}), i.e.
is either $0,1,2$ or $3$. Note that for $n_f=1$ and $n_f = 2$ 
three different algebraisations are available.

If we allow $m$ to be a non-integer and require instead that $\tilde{m}$ with
\begin{equation}
\tilde m \equiv m+bn_f-{1\over 2}n_f 
\end{equation}
is an integer, we conclude that now
\begin{equation} 
\tilde H_N {\cal M}_{\tilde m} \subseteq {\cal M}_{\tilde m} \ .
\end{equation}
In the special case $b=0$,
we can distinguish two different cases:
1) both $m$ and $\tilde{m}$ are integers and 2) only $\tilde{m}$ is an integer. For
1) we find a quadruple algebraisation of the Hamiltonian $\tilde H_N$ (one algebraisation for
$n_f=0$ and three for $n_f=2$):
\begin{subequations}
\begin{equation}
\tilde H_N {\cal M}_m \subseteq {\cal M}_m \  \ {\rm for} \  \ n_f=0 \ , 
\label{a1}
\end{equation}
\begin{equation}
\tilde H_N {\cal M}_{m-1} \subseteq {\cal M}_{m-1} \  \ {\rm for} \  \ n_f=2  \ .
\label{a2}
\end{equation}
\end{subequations} 
Similarly, for 2) we find 
\begin{subequations}
\begin{equation}
\tilde H_N {\cal M}_{m-{1\over 2}} \subseteq {\cal M}_{m-{1\over 2}} \  \ {\rm for} \  \ n_f=1 \ ,
\label{b1}
\end{equation}
\begin{equation}
\tilde H_N {\cal M}_{m-{3\over 2}} \subseteq {\cal M}_{m-{3\over 2}} \  \ {\rm for} \  \ n_f=3 \ .
\label{b2}
\end{equation}
\end{subequations}
Now, $\tilde{m}=m-{1\over 2}$ should be an integer. Again, this is a quadruple
algebraisation of the Hamiltonian $\tilde H_N$ (one algebraisation for
$n_f=3$ and three for $n_f=1$).

\subsection {$a=b=0$~: Relation between the Hamiltonian $H_N$
and the Lam\'e operators } 

In order to understand the pattern of the algebraic solutions
obtained for the model (\ref{oper}), (\ref{potential}), it is useful to study 
the limit $a=b=0$. Using the relation
\begin{equation}
\label{weier}
{\cal P}(x+i\beta) = e_3+(e_2-e_3) {\rm sn}^2(\sqrt{e_1-e_3}x,k)\ \ , \ \ k^2 \equiv {e_2-e_3\over{e_1-e_3}} \ ,
\end{equation}
it is easy to see that for $a=b=0$ the operator (\ref{oper})
takes the form
\begin{equation}
H_N({\bf u}) =(e_1-e_3)\sum_{j=1}^{N} \lbrace 
-{\partial^2\over{\partial u_j^2}} + 2m(2m+1)k^2 
{\rm sn}^2(u_j,k)
\rbrace +2m(2m+1)e_3N 
\label{lame}
\end{equation} 
with $u_j\equiv \sqrt{e_1-e_3}x_j$.
The operator inside the brackets $\{ \}$ of (\ref{lame})
constitutes  $N$ decoupled copies of the Lam\'e operator $L(u)$ :
\begin{equation}
\label{lame1}
L(u) = - \frac{d^2}{du^2} + 2m(2m+1)k^2 {\rm sn}^2(u,k) \ \ \ , \ \ \ 
0 \leq k \leq 1 \ ,
\end{equation}
which 
admits ($4m+1$) algebraic eigenvalues if $m$ is an integer or a half integer.

If $m$ is an integer 
($m+1$) eigenvectors of $L(u)$ are of the form $p_m({\rm sn}^2)$ and 
($3m$) eigenvectors are of the form ${\rm cn}\cdot p_{m-1}({\rm sn}^2)$,
${\rm sn}\cdot p_{m-1}({\rm sn}^2)$, ${\rm dn}\cdot p_{m-1}({\rm sn}^2)$.
sn, cn, dn are abbreviations
for the Jacobi elliptic functions ${\rm sn}(u,k), {\rm cn}(u,k),
{\rm dn}(u,k)$ and $p_n$ denotes a polynomial of degree $n$ in its argument.
If $m$ is a half integer
$3(m+1/2)$ eigenvectors of $L(u)$ are of the form 
${\rm sn\cdot cn}\cdot p_{m+1/2}({\rm sn}^2)$, 
${\rm sn\cdot  dn}\cdot p_{m+1/2}({\rm sn}^2)$,
${\rm cn\cdot dn}\cdot p_{m+1/2}({\rm sn}^2)$ and
($m-1/2$) eigenvectors are of the form 
${\rm sn\cdot cn \cdot dn}\cdot p_{m-1/2}({\rm sn}^2)$.\

Therefore, a total number of $(4m+1)^N$ algebraic eigenvectors
of the Hamiltonian (\ref{lame}) 
can be constructed. However, not all of them
are completely symmetric under the permutations 
of the coordinates.
Since the procedure of algebraisation is crucially related to the
symmetrized variables $\tau_k$ (see (\ref{tau})), only the completely symmetric
solutions can be hoped to be recovered in the generic case
for which $a\neq 0$ and/or $b\neq 0$.\

Studying the solutions of the operator
(\ref{lame}) and the structure of the eigenfunctions
of the Lam\'e operator, it is not difficult to see that the number
of completely symmetric solutions is given by~:
\begin{subequations}
\begin{equation}
C_{m+N}^N + 3C_{m+N-1}^N \  
\end{equation}
if $m$ is an integer and
\begin{equation}
3C_{m'+N}^N + C_{m'+N-1}^N  \ , \   m'\equiv m+\frac{1}{2} 
\end{equation}
\end{subequations}
if $m$ is a half integer,
respectively. $C_q^p$ denotes the usual combinatoric symbol.\

We find that for $b=0$ the number of algebraic solutions
available by applying the method described here agrees nicely with
these above numbers. Moreover, we checked for several
particular cases that, indeed,  the relevant Lam\'e
solutions  are reproduced in the limit $a \rightarrow 0$.
Note that in \cite{gome} only $C_{m+N}^N$ solutions 
were found for integer values of $m$. Our supplementary factorisations
therefore complete the pattern.\

\subsection{The case $N=m=2$, $b=0$} 

For the choice $N=m=2$,
(\ref{a1}) leads to a $6\times 6$ matrix with respect to the basis
$\{1,\tau_1,\tau_2,\tau^2_1,\tau_1\tau_2,\tau_2^2\}$ \cite{gome}:
\begin{equation}
\left( \begin{array}{cccccc}
 0         &g_2(2a+2b+1)&-2ag_3              &4g_3&0  &0 \\
 16a+24b+20&0           &g_2(b+\frac{1}{2})  &4g_2(a+b+1)&2g_3(1-a) &0 \\
 0         &8a+24b+12   &0                   &0&g_2(2a+2b+5) & -4g_3(a+1)\\
 0         &8a+12b+14   &0                   &0&g_2(b+\frac{1}{2}) &2g_3\\
 0         &0           &8a+12b+14           &16(a+3b+3)&0 &g_2(2b+3) \\
 0         &0           &0                   &0&8a+24b+28 &0 \\
\end {array}\right)  \ .
\end{equation}
For (\ref{a2}) we obtain
$3$ different $3\times 3$ matrices with respect to the
 basis $\{1,\tau_1,\tau_2\}$~:
\begin{equation}h_i=
\left( \begin{array}{ccc}
(6 +4 a) e_i & g_2(2a+1)+8e_i^2   & -2 a g_3 \\
14+8a & (10+4a) e_i   & g_2/2+4 e_i^2\\
0   & 28+8a   & (14+4a)e_i  \\
\end {array}\right)  \ , \   i=1,2,3  \ .
\end{equation}
We thus obtain fifteen
algebraic solutions, i.e. an additional nine to the ones obtained
in \cite{gome}.\

In FIGs.~1a and 1b we show the energy eigenvalues as functions of $\epsilon$  for
\begin{equation}
  e_1 = 2 \ \ , \ \ e_2 = -1 + \epsilon \ \ , \ \ e_3 = -1 -\epsilon
\end{equation}
and $a=0$ and $a=5.0$, respectively.
FIG.~1a corresponds to two decoupled Lam\'e operators.
The limit  $\epsilon=0$
further corresponds to the completely integrable case of two
decoupled oscillators ($e_2=e_3$, so $k=0$ and the potential vanishes in 
(\ref{weier}), (\ref{lame})).  
The eigenvalues of this system are of the form
$3(j_1^2 + j_2^2) - 40$ where $j_1, j_2$ are integers. 
The set of algebraic eigenvalues obtained with our factorisation (\ref{factor})
represents just  the completely symmetric
case, i.e. $j_1+j_2=2n$, $n=0, 1, 2,...$ in this limit. This can be checked 
in FIG.~1a. In FIG.~1b the effect of an interaction potential  
on the energy eigenvalues is demonstrated for $a=5.0$.\

The case for which two of the numbers $e_1, e_2, e_3$ are equal
is in itself special, since the fifteen eigenvalues can be 
expressed as linear functions of $a$ and the system is highly
degenerated, irrespectively of $a$. E.g. for $e_1=2$, $e_2=e_3=-1$
\begin{subequations}
three  eigenvalues of the $6\times 6$ matrix are not degenerate~:
\begin{equation}
-8(5+4a) \ , \ -4(7+2a) \ , \   8(1+2a)  \ ,
\end{equation}
the other three eigenvalues of the $6\times 6$ matrix coincide with those of
the $3\times 3$ matrix $h_1$~:
\begin{equation}
-8(2+a) \ ,  \  4(5+4a) \ , \  8(7+2a)  \ ,
\end{equation}
and finally the eigenvalues of the $3\times 3$ matrices $h_2$ and $h_3$
coincide and read~:
\begin{equation}
-2(17+10a) \ , \  -2(5-2a) \ , \  2(7+2a)  \ . 
\end{equation}
\end{subequations}
This is clearly shown in FIG.s~1a and 1b, where at $\epsilon=0$ 
three of the dotted curves, which correspond to three of the eigenvalues of the $6\times 6$ matrix,
and the three dashed curves, which correspond to the three eigenvalues of $h_1$, cross
both for $a=0$ and $a=5.0$, respectively. Similarly, the three solid lines and
the three dotted-dashed lines, which correspond to the three eigenvalues of the
matrices $h_2$ and $h_3$, respectively, cross at $\epsilon=0$.
How these degeneracies disappear for a generic choice of  $e_i$, $i=1,2,3$
is also shown in these figures\

Finally, in FIG.~2 we demonstrate the dependence of the eigenvalues on the parameter $a$
for the special choice 
$e_1=2, e_2=-3/2, e_3=-1/2$.

\section{Summary}
The construction of integrable models of Calogero-Sutherland (CS) type
has recently received a lot of attention in relation to new applications
related to different domains of theoretical physics.
The class of N-body integrable models remains however very tiny and several
generalisations are worth considering.  The construction of quasi exactly
solvable Hamiltonians describing N degrees of freedom appears to be a
possible extension of the notion of integrable systems. As seen in  \cite{gome,inom}
the potential can be more general than those related to the 
root system of a Lie algebra
(typically of the type $A_N$ for potentials depending on 
the differences of the particles' coordinates).

In this paper, we reconsidered such a QES model
proposed recently in \cite{gome}. It depends on four parameters:
two coupling constant $a,b$ and the two periods of the Weierstrass
function ${\cal P}$,  parametrized by $g_2,g_3$. 
More popular  models are recovered for special limits of these constants:
an Inozemtsev model for $b=0$, a system of N decoupled
Lam\'e equations if $a=b=0$ 
and a system of N decoupled oscillators if, in addition,
$e_2 = e_3$ (or equivalently $g_2^3 = 27 g_3^2$). 
We have seen that the case $b=0$ possesses a particularly rich
algebraic spectrum.

By investigation of the spectrum available in these limits, it appears
that the solutions constructed in \cite{gome} do not constitute
the full set of completely symmetric algebraic eigenfunctions of the initial
Hamiltonian  (\ref{oper}).
Following closely
the construction of the Lam\'e polynomials we have found additional algebraisations of the
operator $H_N$.  The set of algebraic eigenfunctions 
obtained in this way coincides exactly with the number of possible
algebraic functions. 
We assume that an extension of the type  of Hamiltonian considered here 
 to $2\times 2$ matrix valued 
operators \cite{bhy} might be possible, but leave this construction as a future
project \cite{bh2}.\\
\\
{\bf Note Added}\\
After the paper was finished several papers appeared dealing with the same topic.
These are e.g. {\it K. Takemura~:  math.QA/0205274} and {\it O. Chalykh et al.~: math.QA/0212029}.\\
\\
\\
{\bf Acknowledgements}\\
Y. B. acknowledges discussions with N. Nachez. B. H. was supported by an EPSRC grant.

\newpage
\begin{fixy}{0}
\begin{figure}
\centering
\epsfysize=10cm
\mbox{\epsffile{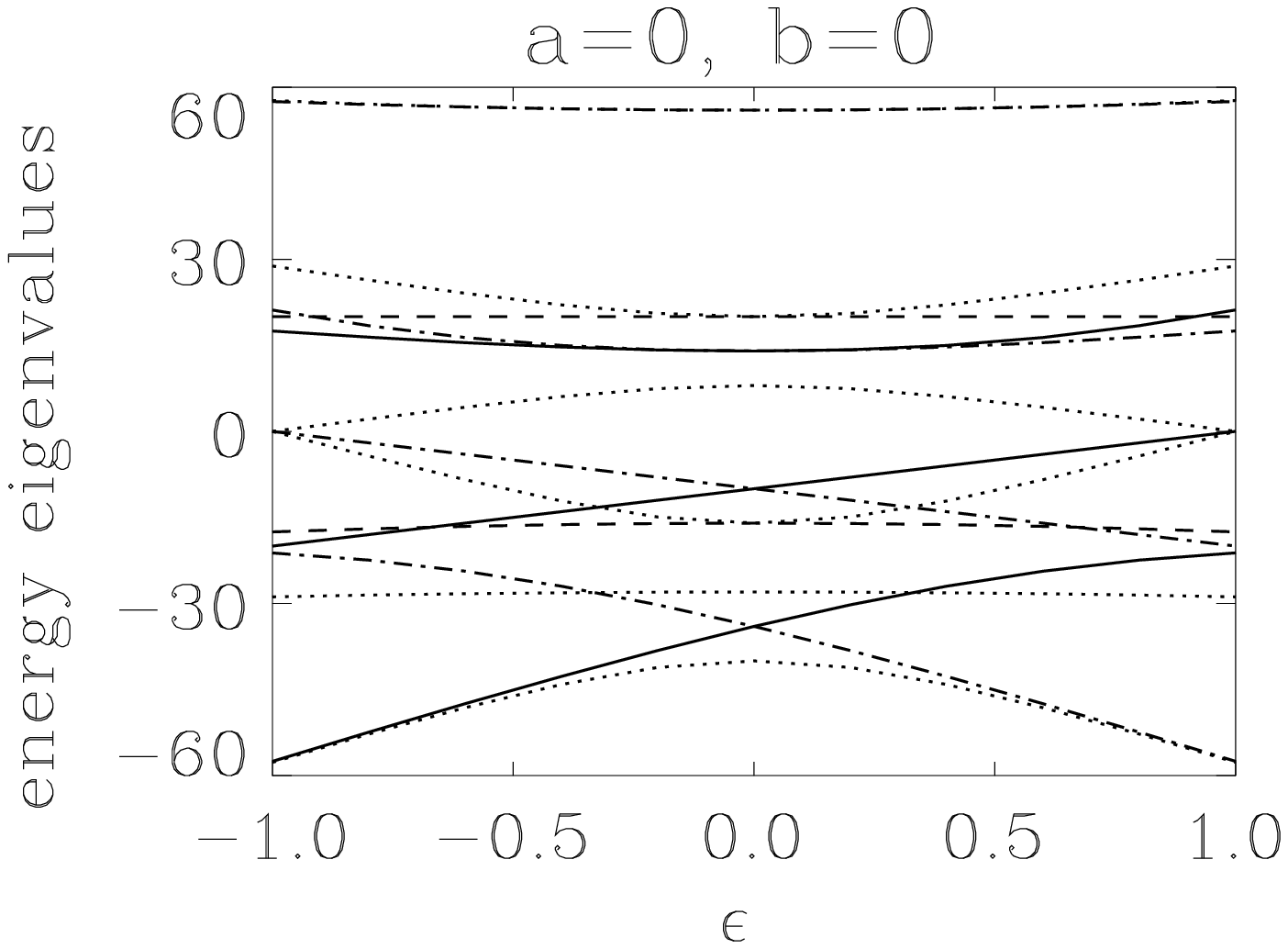}}
\caption{\label{Fig.1a}The energy eigenvalues of the $6\times 6$ matrix (dotted)
and of the $3\times 3$ matrices $h_i$, ($i=1$ dashed, $i=2$ solid, 
$i=3$ dotted-dashed), which correspond to the choice $N=m=2$, are shown
for $a=b=0$ as a function of $\epsilon$, where $e_1=2$, $e_2=-1+\epsilon$
and $e_3=-1-\epsilon$. }
\end{figure}

\begin{figure}
\centering
\epsfysize=10cm
\mbox{\epsffile{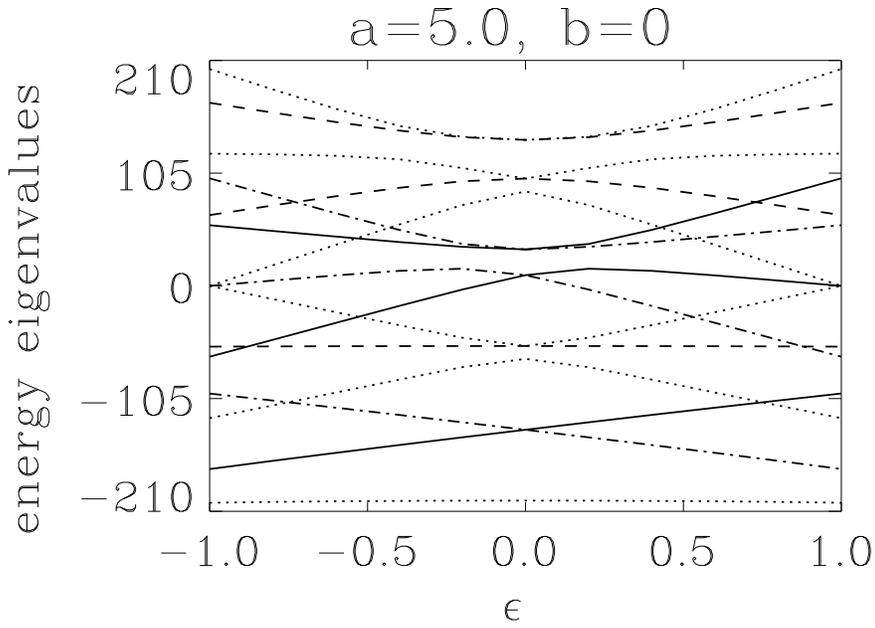}}
\caption{\label{Fig.1b} Same as Fig.1a, but for $a=5.0$. }
\end{figure}
\end{fixy}
\newpage
\begin{fixy}{-1}
\begin{figure}
\centering
\epsfysize=10cm
\mbox{\epsffile{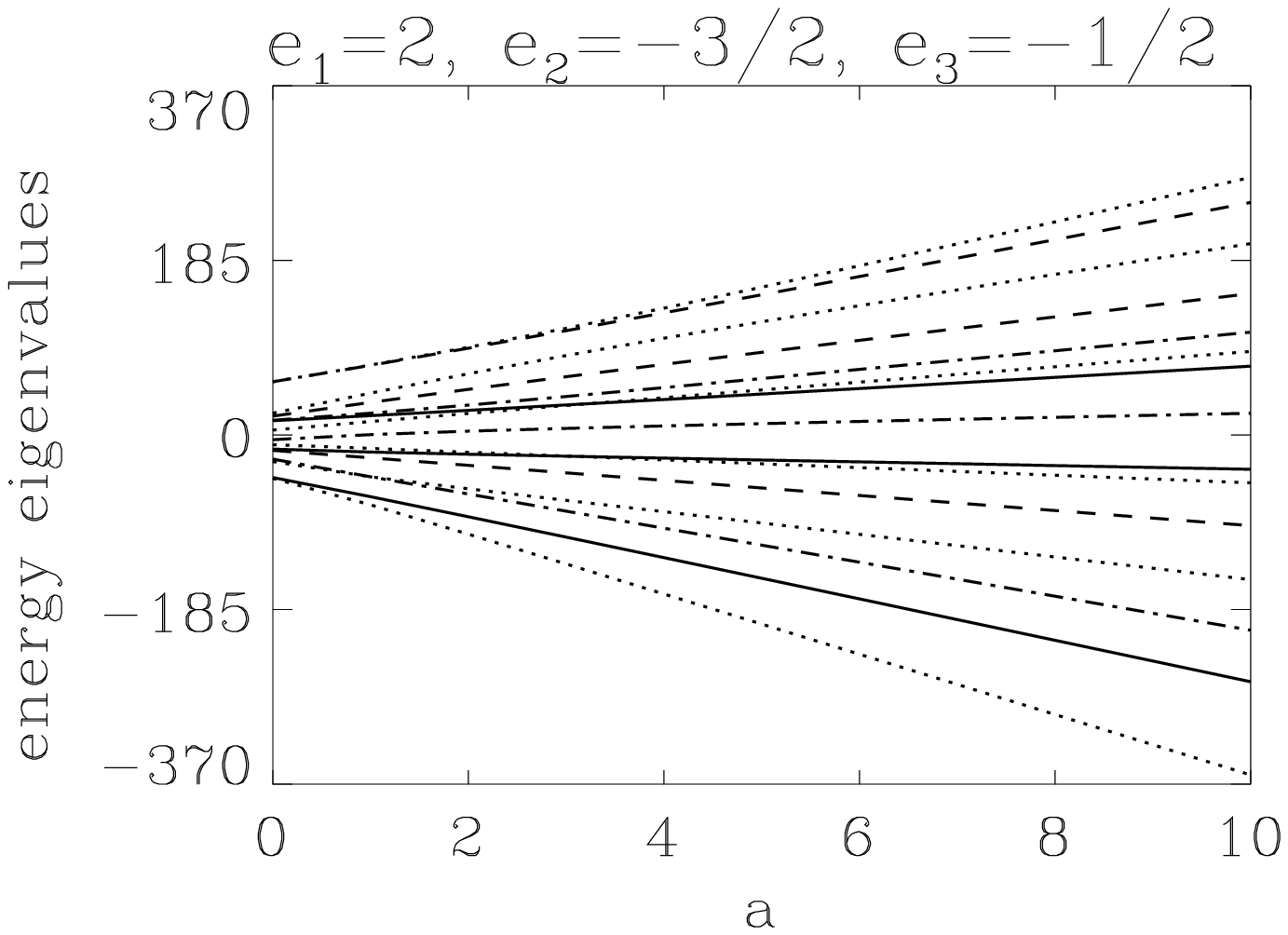}}
\caption{\label{Fig.2}The energy eigenvalues of the $6\times 6$ matrix (dotted)
and of the $3\times 3$ matrices $h_i$, ($i=1$ dashed, $i=2$ solid, 
$i=3$ dotted-dashed), which correspond to the choice $N=m=2$, are shown
for $b=0$ and $e_1=2$, $e_2=-3/2$, $e_3=-1/3$
as a function of $a$. }
\end{figure}
\end{fixy}

\end{document}